# Silicon-Chip Mid-Infrared Frequency Comb Generation

Austin G. Griffith[1], Ryan K. W. Lau[2], Jaime Cardenas[1], Yoshitomo Okawachi[2], Aseema Mohanty[1], Romy Fain[1], Yoon Ho Daniel Lee[1], Mengjie Yu[2], Christopher T. Phare[1], Carl B. Poitras[1], Alexander L. Gaeta[2,3], Michal Lipson[1,3,*]

[1]*School of Electrical and Computer Engineering, Cornell University, Ithaca, NY 14850, USA*

[2]*School of Applied and Engineering Physics, Cornell University, Ithaca, NY 14853, USA*

[3]*Kavli Institute at Cornell for Nanoscale Science, Cornell University, Ithaca, NY 14853 USA*

*\*Corresponding author: ml292@cornell.edu*

Optical frequency combs represent a revolutionary technology for high precision spectroscopy due to their narrow linewidths and precise frequency spacing. Generation of such combs in the mid-infrared (IR) spectral region (2-20 µm) is of great interest due to the presence of a large number of gas absorption lines in this wavelength regime[1]. Recently, frequency combs have been demonstrated in the MIR in several platforms, including fiber combs, mode-locked lasers, optical parametric oscillators, and quantum cascade lasers[2–5]. However, these platforms are either relatively bulky or challenging to integrate on-chip. An alternative approach using parametric mixing in microresonators is highly promising since the platform is extremely compact[6–14] and can operate with relatively low powers. However, material and dispersion engineering limitations have prevented the realization of a microresonator comb source past 2.55 µm[13]. Although silicon could in principle provide a

**CMOS compatible platform for on-chip comb generation deep into the mid-IR, to date, silicon's linear and nonlinear losses have prevented the realization of a microresonator-based comb source. Here we overcome these limitations and realize a broadband frequency comb spanning from 2.1 μm to 3.5 μm and demonstrate its viability as a spectroscopic sensing platform. Such a platform is compact and robust and offers the potential to be versatile and durable for use outside the laboratory environment for applications such as real-time monitoring of atmospheric gas conditions.**

Recent demonstrations of mid-IR frequency comb platforms have enabled excellent sources for spectroscopy. However, the realization of a robust on-chip integrated mid-IR comb source has proven elusive. On-chip integration and miniaturization of the mid-IR comb is critical, as it will enable high portability for stand-off atmospheric sensing out in the field, and well as monolithic integration with other necessary components, such as resonant cavities for gas sensing and photodetectors for measurements. In particular, a CMOS compatible integrated mid-IR comb source would be inexpensive and straightforward for mass production. Mid-IR combs have been previously demonstrated in platforms such as fiber lasers, mode-locked lasers, and optical parametric oscillators[2–4], but these platforms are relatively bulky and cannot be integrated. Supercontinuum generation represents another means for generating broad spectrum in the mid-IR, and has been realized in a number of platforms including silicon waveguides[15]; however, this requires a high peak power pulsed femtosecond source that can generate a broadband coherent spectrum, and for many applications it is desirable to have comb spacings much larger than the ~100-MHz spacing typically produced by such lasers. Another route to mid-IR comb generation is through the use of quantum cascade lasers (QCLs)[5], but the active materials used make on-chip integration difficult.

Microresonator-based combs are promising because they can generate a broad frequency comb in a compact and robust integrated platform. However, material and dispersion engineering limits have prevented the realization of a microresonator comb source past 2.55 μm[13]. Optical frequency combs are a coherent light source consisting of discrete lines that are equally spaced in frequency. Mid-IR comb sources have proved promising for spectroscopy since their broad bandwidth and narrow frequency linewidths make them ideal for probing narrow molecular transitions. With a properly phase-matched geometry, a frequency comb can be generated with a high quality factor microresonator using a single continuous wave (cw) pump laser[6]. Using the parametric $\chi^{(3)}$ nonlinear process of four-wave mixing (FWM), energy is transferred from the pump laser into frequency sidebands. Comb lines will be generated at modes supported by the microresonator and lead to an optical frequency comb with a spacing equal to that of the free spectral spacing of the resonant cavity. Microresonator-based frequency comb sources have been demonstrated in a number of platforms, including silica, quartz, fluoride glasses, silicon nitride, Hydex glass, aluminum nitride, and diamond[6–14]. Parametric four-wave mixing and parametric oscillation require a high quality factor resonator and proper dispersion engineering, While such dispersion engineering has allowed for broad-band combs pumped at 1-μm[16], which has limited the reach of microresonator combs into the mid-IR. In particular, on-chip mid-IR comb generation has not been realized due to the difficulty of creating a highly confined and high-quality microresonator in semiconductor thin films.

Although silicon could provide a CMOS-compatible platform for on-chip comb generation deep into the mid-IR, silicon's linear and nonlinear losses have until now prevented the realization of a silicon microresonator-based comb source. Silicon has a wide transparency window from 1.2 μm to past 8 μm[17] and a large third-order optical nonlinearity ($n_2 = 10^{-14}$ cm$^2$/W at 2.5 μm wavelength[18]), which makes it an excellent platform for mid-IR nonlinear optics. In etched silicon microresonators, quality factors have been limited by scattering losses due to roughness in the waveguide sidewalls, which is made worse by the high index contrast between waveguide core and cladding. The dominant nonlinear loss in silicon in the 2.2-3.3 μm region is three photon absorption (3PA) – a process where three photons are simultaneously absorbed to excite an electron-hole pair[19]. The number of photons lost directly to 3PA are small (dominated by linear waveguide losses), but the generated free-carrier population will induce significant optical losses[20].

In order to overcome silicon's linear losses, we use a novel thermal oxidation based "etchless" fabrication process to achieve a high quality factor of 590,000 in a silicon microresonator at a wavelength of 2.6 μm. The silicon waveguide is defined in an etchless manner using thermal oxidation, instead of dry etching, to form the waveguide core[21,22]. We deposit low pressure chemical vapor deposition silicon nitride on top of a 500 nm silicon-on-insulator (SOI) wafer. The silicon nitride serves as an oxidation mask. We then pattern and etch only the silicon nitride, then thermally oxidize the wafer. The silicon underneath the nitride mask doesn't oxidize, leaving the silicon waveguide core intact. This method avoids the roughness and absorption sites that can be introduced by reactive ion etching[23]. See methods for full fabrication process. The result is a 590,000 intrinsic quality factor resonator at 2.6 μm, which to our knowledge, is the highest yet achieved for a silicon resonator in the mid-IR. This corresponds to a propagation loss of 0.7 dB/cm. The geometry of the silicon waveguide governs the bandwidth of the frequency comb since the waveguide cross-section determines its dispersion profile[8,24].

In order to mitigate silicon's nonlinear loss, we embed the resonator in a reverse biased PIN diode junction to sweep out carriers generated from three-photon absorption. Significant free carriers accumulate when pumping a passive silicon waveguide with a cw laser, limited only by the natural free-carrier lifetime of the structure. Here we counteract the carrier generation while using a cw pump is to extract carriers using a PIN junction operated in reverse bias[25]. The PIN junction prevents the electrical injection of carriers into the waveguide while allowing generated free carriers to be swept out – with effective free carrier lifetimes demonstrated as short as 12 ps[26].

We demonstrate the generation of an optical frequency comb between 2.1 μm and 3.5 μm under reverse bias conditions. In order to generate the frequency comb shown in Fig. 2, we coupled 150 mW of optical power into the bus waveguide composed of a 1.4-μm wide etchless waveguide and tuned the frequency of the pump to be on-resonance with the microresonator. We extract 2.7 mA of current with 10 volts applied to the PIN junction. This relatively high current is an indication that indeed that the three-photon absorption is significant in these structures and therefore the reverse PIN geometry is crucial for achieving oscillations. The generated frequency comb has a line spacing of 130 GHz, corresponding to the FSR of the microresonator. We measure an oscillation threshold with a low power of 3.1 +/- 0.6 mW in the bus waveguide. In order to resolve the full extent of the frequency comb and overcome the limited dynamic range for the optical spectrum analyzer (OSA), we filter the output lines using a series of optical bandpass filters. These filters attenuate the pump and the central part of the comb, allowing us to reach the noise floor of the OSA. The comb spectrum as shown in Fig. 2 is then corrected for the spectral profile of the filters.

We observe a strong influence of the comb spectral shape on the voltage applied to the diode, indicating that indeed the control of the carrier population through the PIN diode is critical for achieving the wide combs. With the voltage source off, combs are generated, however their bandwidth is limited by the natural free-carrier lifetime of the waveguide – which was measured to be 5.23 ns +/- 0.83 ns using a pump-probe experiment. See methods for details. Although a few comb lines can be generated with the PIN turned off, the PIN is required to achieve broadband comb generation above the experiment noise floor – indicated by the dashed line shown in Fig. 3. By increasing the voltage, we find that we can control the comb shape by dynamically changing the carrier lifetime in the ring. In order to simulate the spectral-temporal dynamics of the microresonator combs, we use a recently developed numerical approach[24,27] based on the Lugiato-Lefever equation[28,29]. We show the simulated comb spectra for various free-carrier lifetime values in the right column of Fig. 3 and confirm that a reduced free-carrier lifetime results in broadband comb generation.

In order to demonstrate the feasibility of this comb as a source for spectroscopy, we show proof-of-concept gas sensing in the mid-IR. Since our comb source generates lines beyond 3 μm, we have access to a significant number of molecular gases that absorb in that range. Acetylene ($C_2H_2$) has strong vibrational absorption in the 3.0 to 3.1 um range, and consists of a set of comb-like absorption lines as shown in Fig. 4A (retrieved from the HITRAN database[30]). We first filtered the comb to keep only those lines near the acetylene absorption band. We measured the transmission of the comb through an air-filled gas cell, as shown in Fig. 4B using the OSA. We then filled the gas cell with an 8% acetylene/air mixture, and measured the transmission again. Despite the fact that the spacing of the acetylene absorption peaks and the comb lines are different, we were able to measure strong attenuation corresponding to four of the $C_2H_2$ absorption lines (Fig. 4C). Note that since we are using an integrated microresonator, the line spacing can in principle be engineered independently of the dispersion profile, to enable detection of numerous spectroscopic lines and therefore high sensitivity detection of a particular gas.

In summary we have demonstrated the first on-chip integrated mid-IR frequency comb source generated in an etchless silicon microresonator platform and the first demonstration of parametric oscillation in a silicon platform. This platform can enable a versatile and straightforward source for mid-IR gas spectroscopy and opens the door for compact and robust mid-IR gas sensors. The ability to achieve comb generation in an integrated platform will enable practical realization of completely integrated mid-IR comb based spectrometers usable in a myriad of environments.

**Methods:**

**Device fabrication.** A thermal oxidation based "etchless" fabrication process was used to fabricate the devices. Using a commercial silicon-on-insulator wafer with a 500 nm silicon layer and a 3 µm buried oxide, 200 nm of silicon nitride was deposited using low-pressure chemical vapor deposition (LPCVD). MaN-2405 electron-beam resist was used to pattern the nitride layer. After exposure and development, the silicon nitride was etched in fluorine chemistry. The wafer was then thermally oxidized at 1200 °C to form the etchless waveguide structure. The wafer was oxidized for 20, 50, and 60 minutes of dry, wet, and dry oxidation respectively.  The slab in the taper region was then etched away. The wafer was then clad with 2 μm of PECVD silicon dioxide. The oxide was then etched away next to the waveguides, and then the slab was doped with boron and phosphorous on either side to form a PIN diode. Finally, the wafer was metalized, and wires were patterned and etched.

**Free-carrier lifetime measurement.** For carrier lifetime measurements, we perform pump-probe experiments with a counter-propagating pulsed pump and CW probe.  An Erbium-doped fiber laser (1550nm, 500fs FWHM) is amplified to a peak power of ~200W in an EDFA, then coupled via a circulator into one facet of the chip.  A CW probe (1575nm) is coupled into the opposite facet and appears on the third port of the circulator, isolated from the pump.  Neither laser overlaps with a ring resonance; we measure only the bus waveguide.  We further improve probe isolation by wavelength filtering the third port output with a grating filter, then route this fiber to an optical sampling oscilloscope.  The oscilloscope is triggered off the pump laser's fast sampling photodiode, giving us a time-domain measurement of the transmission of the waveguide after strong pump excitation.  Absorption increases sharply when the pump-generated free carriers flood the waveguide, then exponentially decreases as they recombine.  We take the time constant of this exponential to be the free-carrier lifetime.

**Microresonator comb simulation.** To model the microresonator comb, we use the LL equation, modified to include contributions from higher-order dispersion, self-steepening, multi-photon absorption, and free-carriers. Free-carrier generation is governed by three-photon absorption, and recombination is modeled with a single exponential with the effective lifetime as the time constant. Full equations and parameter values can be found in Ref. 27.

## Acknowledgements

The devices were fabricated at the Cornell Nano-Scale Science & Technology Facility. The authors gratefully acknowledge support from DARPA for award # W31P4Q-13-1-0016. The authors gratefully acknowledge support from AFOSR for award # BAA-AFOSR-2012-02 supervised by Dr. Enrique Parra. This work made use of the Cornell Center for Materials Research Shared Facilities which are supported through the NSF MRSEC program (DMR-1120296). Austin G. Griffith acknowledges the National Defense Science and Engineering Graduate Fellowship (NDSEG) for funding.

## Author Contributions

A.G.G. designed and fabricated the devices, and performed the experiment. A.G.G. analyzed the data. R.K.W.L. performed the comb simulations. J.C. assisted in designing the devices. J.C., A.M., and C.B.P. assisted in fabricating the devices. Y.O., R.K.W.L., M.J., and R.F. assisted in the experiments. Y.H.D.L. and C.T.P. performed the carrier lifetime measurement. J.C., Y.O., R.K.W.L., A.L.G. and M.L. contributed to data analysis. A.L.G. and M.L. supervised the project. All authors discussed the results and commented on the manuscript.

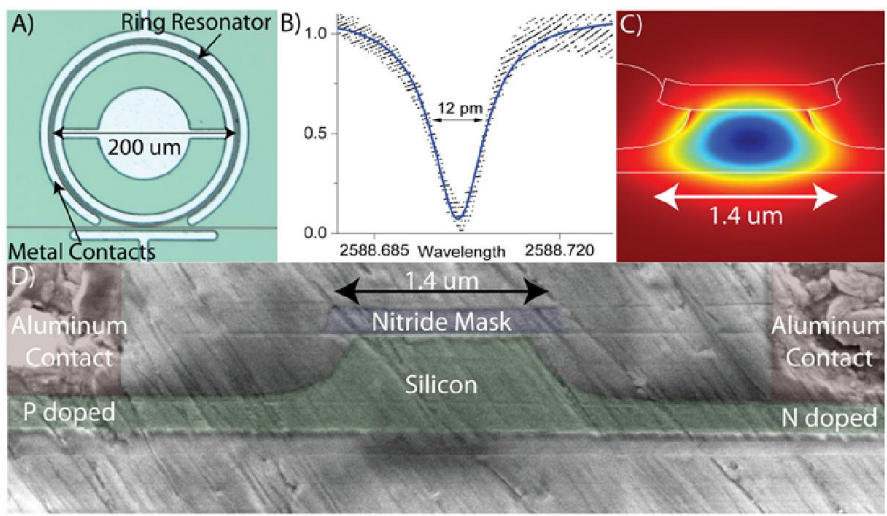

Figure 1| 1(A) Optical microscope image of a ring resonator and metal contacts fabricated using the etchless process. 1(B) Normalized transmission spectrum taken at low input power, showing a resonance with Q=590,000. 1(C) Simulated optical mode at 2.6 um, showing high modal confinement in Si. 1(D) False colored cross-sectional SEM image of silicon waveguide, doped regions, and metal contacts.

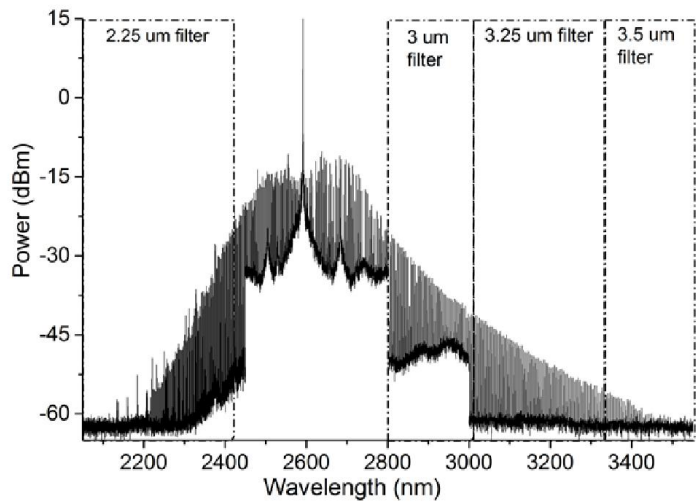

Figure 2| Broadband frequency comb generation from 2.1 μm and 3.5 μm in the etchless silicon micro-resonator. This frequency comb is generated with 150 mW of optical power in the bus waveguide, and with a 10 V reverse bias applied on the PIN junction. The frequency spacing of the comb is 130 GHz. Due to the limited dynamic range of the optical spectrum analyzer, the frequency comb was measured using a series of optical filters.

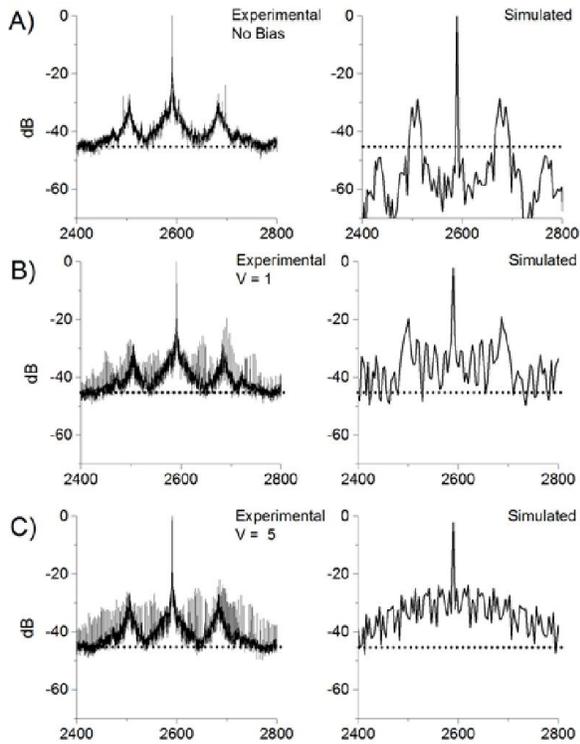

Figure 3 | Rows A-C: The effect of three photon absorption on frequency comb generation depends strongly on the bias voltage of the PIN structure, with a dashed line showing the approximate noise floor. (A) With the pump set at a fixed wavelength and the voltage source off, only a few lines are generated. (B) By applying voltage in reverse bias to the junction, we observe dramatically more comb lines in both experiment (left) and simulation (right). (C) At high voltages, we achieve broad comb generation.

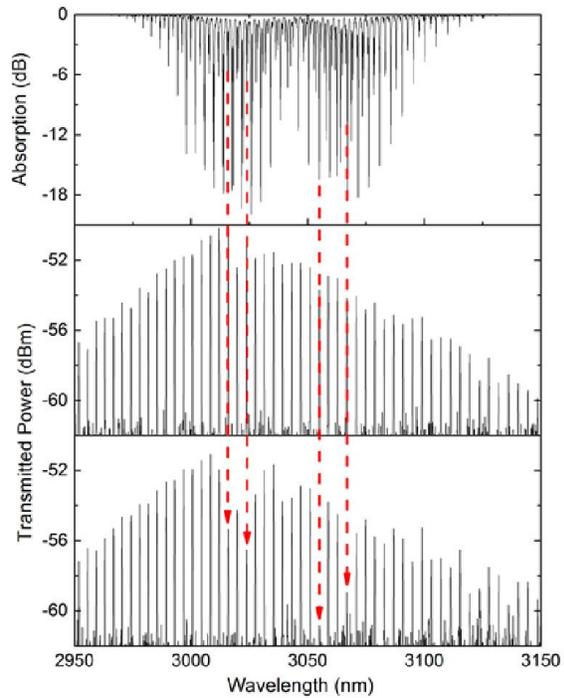

Figure 4| Demonstration of mid-IR gas sensing using frequency comb source. (A) Absorption spectrum of acetylene gas near 3.0 μm from the HITRAN database[30]. (B) Portion of the frequency comb transmitted through empty gas cell. (C) Transmission of the frequency comb when the gas cell contains 8% Acetylene/Air mixture. One can see the strong attenuation of four comb lines, corresponding to four of the acetylene's absorption lines.